# Iridium-doping as a strategy to realize visible light absorption and p-type behavior in BaTiO$_3$

*Sujana Chandrappa, Simon Joyson Galbao, P. S. Sankara Rama Krishnan, Namitha Anna Koshi, Srewashi Das, Stephen Nagaraju Myakala, Seung-Cheol Lee, Arnab Dutta, Alexey Cherevan, Satadeep Bhattacharjee and Dharmapura H. K. Murthy[*]*

Sujana Chandrappa, Simon Joyson Galbao, Dharmapura H. K. Murthy
Department of Chemistry, Manipal Institute of Technology, Manipal Academy of Higher Education, Manipal, Karnataka, 576104, India.
E-mail: murthy.dharmapura@manipal.edu

Dharmapura H. K. Murthy
Center for renewable energy
Manipal Institute of Technology, Manipal Academy of Higher Education,
Manipal, Karnataka, 576104, India.

P. S. Sankara Rama Krishnan
School of Materials Science and Engineering, Nanyang Technological University, 50 Nanyang Avenue 639798, Singapore.

Namitha Anna Koshi, Seung-Cheol Lee, Satadeep Bhattacharjee
Indo-Korea Science and Technology Center (IKST), Jakkur, Bengaluru 560065, India.

Srewashi Das, Arnab Dutta
Department of Chemistry, Indian Institute of Technology Bombay, Powai, Mumbai 400076, India.

Stephen Nagaraju Myakala, Alexey Cherevan
Institute of Materials Chemistry, Technische Universität Wien, 1060, Vienna, Austria.

**Abstract**

BaTiO$_3$ (BTO) is typically a strong n-type material with tuneable optoelectronic properties via doping and controlling the synthesis conditions. It has a wide band gap that can only harness the ultraviolet (≤ 390 nm) region of the solar spectrum. Despite significant progress, achieving visible-light absorbing BTO with tuneable carrier concentration has been challenging, a crucial requirement for many applications. In this work, a p-type BTO with visible-light (λ≤600 nm) absorption is realized via iridium (Ir) doping. Detailed analysis using advanced spectroscopy tools and computational electronic structure analysis is used to rationalize the n- to p-type transition after Ir doping. Results offered mechanistic insight into the interplay between the dopant's site occupancy, the dopant's position within the band gap, and the defect chemistry affecting the carrier concentration. A decrease in the Ti$^{3+}$ donor levels concentration and the mutually correlated oxygen vacancies upon Ir doping is attributed to the p-type behavior. Due to the formation of Ir$^{3+}$/Ir$^{4+}$ in-gap energy levels within the forbidden region, the optical transition can be elicited from or to such levels resulting in visible-light absorption. This newly developed Ir-doped BTO can be a promising p-type perovskite-oxide with imminent applications in solar fuel generation, spintronics and optoelectronics.

## 1. Introduction

Owing to their tunable properties, ATiO$_3$-type oxide perovskites (A= Ba, Sr, Ca) are promising materials with a range of applications in optoelectronics,[1] gas sensors,[2] transistors,[3] catalysis,[4,5] batteries,[6] and photocatalysis.[7–9] Importantly, their performance is determined by their carrier concentration (n- or p-type), which depends on the ionic vacancies, doping, and defect chemistry/density.[10,11] BaTiO$_3$ (BTO), one of the widely known perovskite oxides, typically exhibits a strong n-type behavior (carrier concentration ≥10$^{19}$ cm$^3$).[12] Due to the size differences of Ba and Ti, BTO is highly susceptible to substitution/doping of various metal ions at both sites, thus offering a key tool to control its carrier concentration. In addition, the synthesis environment (reducing or oxidizing), a ratio of the precursors, and oxygen partial pressure significantly impact the carrier concentration and its optoelectronic properties.[13–16]

In the ideal stoichiometric BaTiO$_{3.0}$, the oxygen content and the total amount of metallic constituents (Ba$^{2+}$ and Ti$^{4+}$) are chemically equivalent. Hence, BTO is expected to be an intrinsic semiconductor. However, owing to ubiquitous defects, as a result of synthesis conditions and/or (unintentional) doping, BTO is inherently non-stoichiometric (BaTiO$_{3\pm x}$). The defect equilibrium that renders a strong n-type behavior to BTO can be understood with the help of Equations 1 and 2. Under thermal equilibrium, owing to the inadequate oxygen, some Ti-O bonds (lattice oxygen) dissociation takes place. This process releases molecular oxygen leading to the formation of non-stoichiometric $BaTiO_{3-x}$, concurrently giving rise to oxygen vacancies as depicted in equation 1. Additionally, the oxygen vacancies are accompanied by free electrons, which are released into the host lattice (refer to Equation 1) and can reduce $Ti^{4+}$ ($3d^0$) to $Ti^{3+}$ ($3d^1$), as indicated by Equation 2. Such $Ti^{3+}$ species are typically located close to the conduction band (CB) and act as donor levels,[17,18] which effectively increases the carrier concentration. Therefore, the presence of $Ti^{3+}$ levels in oxygen-deficient, non-stoichiometric $BaTiO_{3-x}$, renders strong n-type behavior.[19]

$$BaTiO_3 = BaTiO_{3-x} + \frac{1}{2}xO_2 + 2xe^- + xV_0 \quad (1)$$

$$BaTiO_3 = BaTi(VI)_{1-2x}Ti(III)_{2x}O_{3-x} + \frac{1}{2}xO_2 + xV_0 \quad (2)$$

Among its many applications, BTO is also being used for solar fuel generation via photocatalysis. In this process, by virtue of its semiconducting behavior, charge carriers generated upon light absorption are used for reduction and oxidation reactions.[20] BTO has a wide optical band gap of ≈3.2 eV, is capable of absorbing the ultraviolet (UV) part of the solar spectrum. It is extensively utilized for photocatalytic solar fuel production, like its analogue, $SrTiO_3$.[21,22] Besides being chemically stable in aqueous conditions, the energetic positions of CB and VB offer thermodynamic driving energy to drive both reduction and oxidation of water. Another unique advantage of BTO is its ferroelectric behavior (in the tetragonal phase) that can be harnessed to enhance the charge photogeneration yield.[23,24] Recently, piezo-photocatalysis [25–27] has gained significant attention, wherein BTO-based materials hold promise towards cost-effective solar fuel production under ambient conditions.

The importance of carrier concentration in a photocatalyst (n- or p-type) on solar fuel efficiency has recently been established. Aluminium-doped $SrTiO_3$ photocatalyst has recently demonstrated ≈0.8% solar-to-hydrogen (STH) efficiency and remarkable stability, surpassing 200 days in 100 $m^2$ photocatalyst modules.[28] Note that to realize a high STH efficiency, $Al^{3+}$ doping of $SrTiO_3$ was indispensable.[29] Interestingly, Al-doping rendered p-type behavior to $SrTiO_3$ owing to $Al^{3+}$ doping at $Ti^{4+}$-site of $SrTiO_3$, consequently decreasing the $Ti^{3+}$ donor levels that contribute to n-type behavior in $SrTiO_3$.[30,31] Thus, realizing a controlled transition between n- and p-type nature is a promising strategy to enhance the photocatalytic $H_2$ evolution efficiency in oxide-based perovskites.

In the example discussed above, $Al^{3+}$ dopant did not form new defect levels within the forbidden region of the $SrTiO_3$ band gap. Thus, Al-doped $SrTiO_3$ retained virtually the same wide band gap (≈3.2 eV) as the undoped, absorbing light exclusively in the UV region that constitutes <5% of the solar spectrum. Besides inducing p-type behavior in photocatalysts, red-shifting the optical absorption to harness the visible region of the solar spectrum is essential. In this direction, very few reports show p-type behavior with simultaneous band-gap narrowing via doping rhodium at $Ti^{4+}$-sites of both $SrTiO_3$ and BTO.[32–34] Another important prospect of visible-light absorbing p-type materials is to utilize them as photocathodes in z-scheme in conjunction with n-type photoanode materials.[35–37] Rh-doped $SrTiO_3$ is employed recently with molybdenum-doped $BiVO_4$ to achieve a remarkably high 1% STH efficiency.[36] Thus, there is a need for p-type photocatalysts that show promising photocatalytic activity under visible/near infrared light.

Despite extensive research and prospects of BTO, only a very few visible-light absorbing p-type BTO are reported. Realizing visible-light absorption and a p-type BTO is not a straightforward process. This process demands a comprehensive understanding of the interplay between the number of mutually correlated and complex parameters, such as (i) the choice of dopant, (ii) site occupancy of dopant (Ba/Ti-site) within the host lattice, (iii) the energetic position of the dopant energy levels within the forbidden region and its electron occupancy, and (iv) change in the electronic nature of the defects due to doping. All these parameters collectively determine the carrier concentration and optical absorption properties. However, it has been challenging to understand the factors involved in realizing p-type and visible-light-absorbing BTO. There is an immense scope to experimentally synthesize such material for a wide range of applications.

In this work, we realized n- to p-type transition in BTO and extended its absorption towards the visible region ($\lambda$ ≤600 nm) through Ir doping at Ti sites. Doping-induced p-type behavior is rationalized based on the changes in the structural, surface, and optoelectronic properties using a range of complementary spectroscopy techniques and computational analysis of the electronic structure. A detailed mechanistic insight into the defect chemistry involved in p-type behavior is also provided. To the best of our knowledge, using Ir as a dopant to achieve p-type BTO is the first of its kind. This material will find promising applications in optoelectronics and photocatalysis.

## 2. Results and discussion

Doping-induced changes in carrier concentration reflect a change in the structural, surface, and optoelectronic properties and are strongly dependent on the site of occupancy of the Ir in BaTiO$_3$. In this direction, probing the site of occupancy (Ti$^{4+}$ site or Ba$^{2+}$ site) of Ir is essential and will constitute the first part of the results and discussion. This part would aid in explaining the origin of a visible-light absorption and p-type behavior upon Ir-doping, which will be examined in the second part of the results and discussion.

### 2.1. Determining the site of occupancy

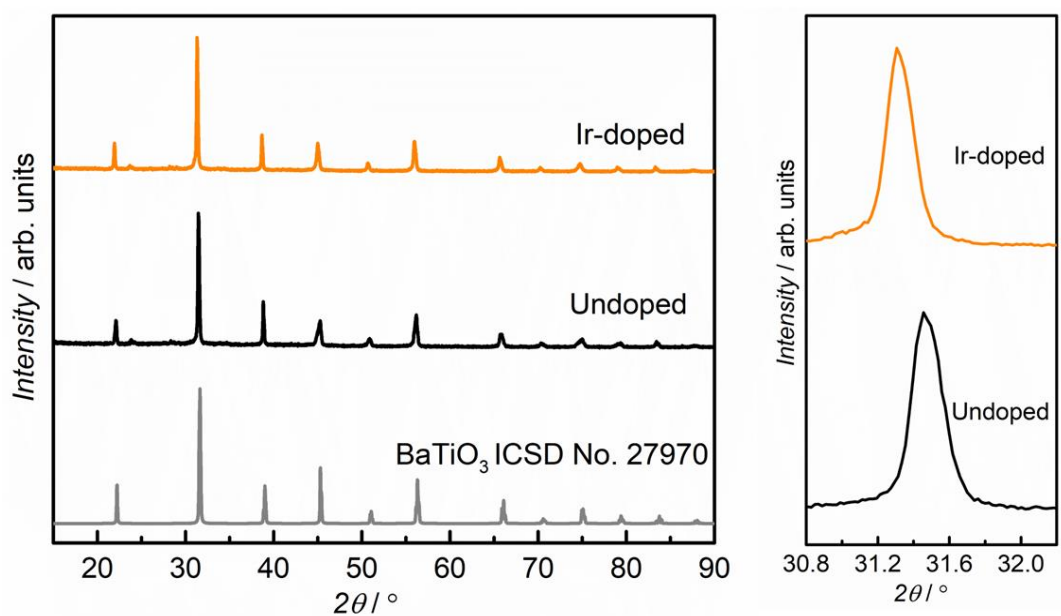

**Figure 1**. The XRD patterns of undoped and Ir-doped BaTiO$_3$ powder samples.

**Figure 1** presents the XRD patterns of the synthesized undoped BTO and Ir-doped BTO. Both samples show peaks characteristic of BTO cubic phase, according to the ICSD database (ICSD No. 27970). Owing to the susceptibility of BTO structure to doping at both Ba-site and Ti-site, determining the site of occupancy of Ir in BTO host lattice is rather difficult. The ionic radii of Ba$^{2+}$, Ti$^{4+}$ and Ir$^{4+}$ is 161 pm, 60.5 pm and 75.6 pm, respectively. Solely based on the ionic radii values, Ir$^{4+}$ is expected to substitute for a bigger sized Ba$^{2+}$ site, which would result in lattice contraction leading to the shift of XRD peaks towards higher two theta values. On the contrary, compared to undoped, a shift of around 0.16 $2\theta°$ is observed towards lower values for Ir-doped BTO hinting lattice expansion. Additionally, the calculated lattice parameter of undoped BTO is 0.401 nm is in agreement with that of undoped BTO reported earlier.[46] After

Ir-doping, the lattice parameter increases to a value of 0.4029 nm, indicating lattice expansion. Both these observations are correlated to lattice expansion which can only be explained by considering that Ir-dopant occupies the $Ti^{4+}$ site. With $Ir^{4+}$ occupying $Ti^{4+}$-site at the body centered cubic (bcc) position in $TiO_6$ octahedral of the BTO lattice, the difference in ionic radii would lead to structural distortion resulting in lattice expansion. Besides, Ir-dopant is expected to occupy the $Ti^{4+}$-site as $Ir^{4+}$ in consideration of charge neutrality. These observations suggest Ir is likely to be doped for $Ti^{4+}$ site.

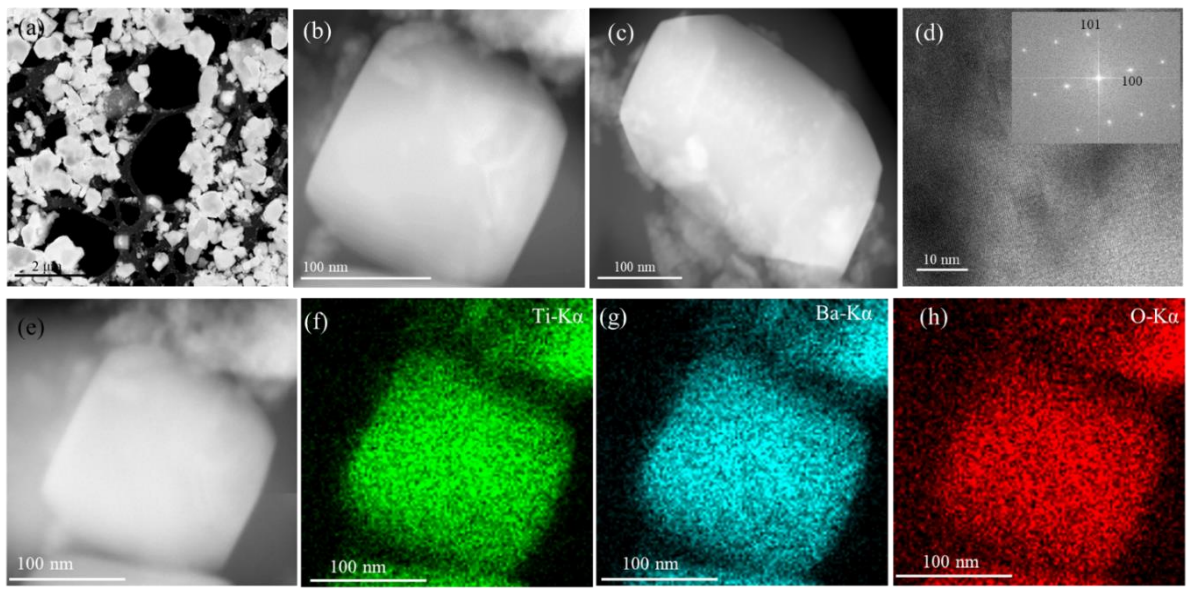

**Figure 2**. a-c) STEM-HAADF images. d) TEM-Bright field image with SAD pattern and FFT inset. e-h) STEM-EDS image and mapping corresponding to Ti (f), Ba (g), O (h). i) EDS spectra of undoped BTO.

**Figure 2** (a-c) shows the STEM-HAADF (high angle annular dark field) images of undoped BTO. In addition to the random distribution of particles, two distinct morphologies of the powder samples are observed. Square particle of ~80 to 100 nm and rectangle particle of size ~200 nm is observed (b-c) and both morphologies show a very uniform bright contrast, confirming the uniform distribution of Ba, Ti and O. TEM-bright field image (Figure 2d) shows uniformly distributed lattice and in the inset Fast Fourier Transform (FFT) confirms the cubic symmetry of undoped BTO, and is in agreement with XRD data in Figure 1. STEM-EDS mapping of Ba, Ti and O in Figure 2 (e-h) shows a very distinct contrast of each element, indicating the uniform distribution of elements.

The influence of Ir doping on the structural modification in the BTO lattice was further analyzed. **Figure 3** (a-b) shows the STEM-HAADF image of Ir-doped BTO and confirms the presence of pristine and Ir-doped regions with dark with bright contrast, respectively.[47,48] The contrast in STEM-HAADF images is observed due to the differences in the atomic number, and bright contrast observed in this sample arises due to the presence of Ir as it is heavier than Ba and Ti. The bright contrast extends over a few unit cells, thereby confirming the distribution of Ir is highly localized, expected for relatively 3 mol % low dopant level. STEM-EDS mapping in Figure S2 (e-h) of Ba, Ti, O and Ir feature distinct contrast of individual elements evidencing the uniform elemental distribution. Thus, ruling out the possibility of Ir segregation or phase separation in the doped sample.

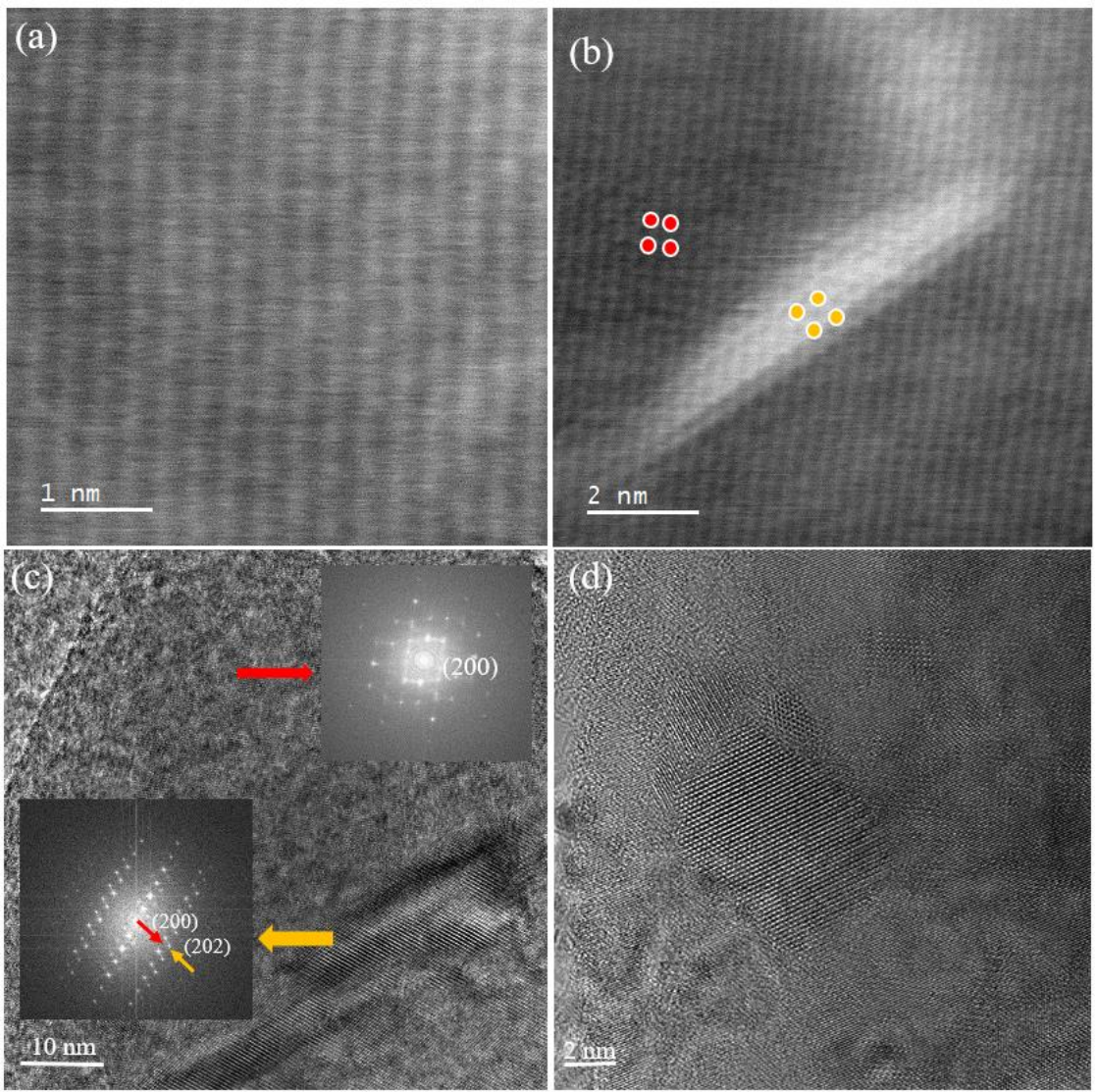

**Figure 3**. STEM-HAADF images Ir-doped BTO showing (a)pristine and (b)doped regions. TEM-bright field images of Ir-doped BTO (c and d), with FFT in the inset (c).

Figure 3b compares dark (red circles) and bright (yellow circles) contrast regions that correspond to pristine and doped regions of Ir-BTO, respectively. Distinctive regions noted after Ir-doping indicates subtle lattice positional changes after doping. To further probe the dopant's site occupancy in the BTO host lattice, STEM-HAADF study was conducted. Note that if Ir substitutes for the Ba-site of BTO, a lattice shrinkage is expected because $Ba^{2+}$ is at least two times bigger ionic radius than $Ir^{4+}$ or $Ir^{3+}$. However, STEM-HAADF data (Figure S2c) indicated an increase in the lattice parameter value by ≈0.05 nm, which can be explained due to lattice expansion. This observation implies the Ir substitution at Ti-sites of the BTO lattice. The HR-TEM images of Ir-doped BTO in Figures 3c and 3d indicate region of pristine BTO lattice and where the lattice is distorted due to Ir doping. The FFT (inset in Figure 3c) from the pristine region (yellow arrow) of the Ir-doped sample relates to the (200) of BTO, while the Ir-doped region matches with (202) orientation (red arrow).[49,50] Such a change in crystal lattice orientation from (200) to (202) plane after doping is correlated to the structural modifications in the $TiO_6$ octahedra after doping, further corroborating Ir doping at Ti sites. Thus, inference from TEM analysis is in good agreement with the shift of two theta degrees towards lower values upon Ir doping, from XRD analysis discussed earlier.

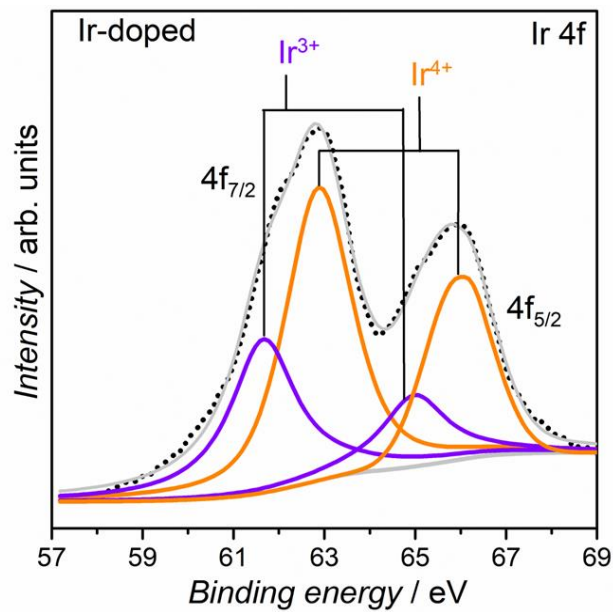

**Figure 4**. High-resolution XPS core level spectrum of Ir-doped BTO showing the Ir 4f region.

To further understand the defect chemistry on doping, surface chemical state of various constituent elements of BTO is investigated by XPS. **Figure 4** depicts the Ir 4f spectra of Ir-doped BTO. Ir 4f spectral deconvolution yielded a full width half maxima maintained (FWHM) of 1.75. The characteristic Ir $4f_{7/2}$ peak is reported to appear between 62.3 eV and 62.8 eV for Ir in 4+ oxidation state and between 61.6 eV and 62.0 eV for Ir in 3+ oxidation state.[51–53] The binding energy of Ir $4f_{7/2}$ at 62.80 eV (indicated by orange) corresponds to $Ir^{4+}$, while that at 61.66 eV (indicated by violet) can be attributed to $Ir^{3+}$. The relative percentage of $Ir^{4+}/Ir^{3+}$ is 1.68, which evidence that Ir-dopant in BTO host exists predominantly as $Ir^{4+}$.

To elucidate the changes in the chemical state(s) of Ti-species on Ir-doping, Ti 2p core level spectra were analyzed in detail. **Figures 5**a and 5b depict Ti 2p spectra of undoped and Ir-doped BTO, respectively. To obtain the best fit for Ti 2p spectra of both samples, deconvoluting the spectra into three components was essential, indicating the presence of Ti in three different oxidation states. **Table 1** lists fitting parameters and individual contributions from $Ti^{4+}$, $Ti^{3+}$, and $Ti^{2+}$ before and after doping. In undoped BTO, observing a characteristic doublet at 458.30 eV ($2p_{3/2}$) and 463.98 eV ($2p_{1/2}$) indicated the presence of $Ti^{4+}$.[54] Besides, pronounced peaks at 457.14 eV ($2p_{3/2}$) and 462.77 eV ($2p_{1/2}$) are attributed to the presence of $Ti^{3+}$, which is in good agreement with earlier reports.[55,56] The presence of $Ti^{2+}$ is evidenced by peaks at 456.73 eV ($2p_{3/2}$) and 461.60 eV ($2p_{1/2}$).[57] For undoped BTO, Ti exists predominantly (70.86 %) in $Ti^{3+}$ states, endorsing a strong n-type behavior.

**Table 1**: Percentage composition of $Ti^{4+}$, $Ti^{3+}$ and $Ti^{2+}$ in undoped and Ir-doped BTO.

| Sample | $Ti^{4+}$ [%] | $Ti^{3+}$ [%] | $Ti^{2+}$ [%] |
|---|---|---|---|
| Undoped BTO (FWHM-1.24) | 11.45 | 70.86 | 17.69 |
| Ir-doped BTO (FWHM-1.20) | 29.65 | 59.46 | 10.89 |

Next, Ti 2p core level spectra of Ir-doped BTO (Figure 5b) was analyzed. Clearly, $Ti^{4+}$ signal intensity increased from 11.45 % in undoped BTO to 29.65 % in Ir-doped BTO, as shown in Table 1. The increase of $Ti^{4+}$ signal on Ir-doping is accompanied by a mutual decrease in the signal intensity of lower valent Ti-states ($Ti^{3+}$ and $Ti^{2+}$), compared to undoped BTO. The relative concentration of the Ti species listed in Table 1 evidenced a factor of two point six times increment in the $Ti^{4+}$ concentration upon Ir-doping, consequently a decrease in the lower-valent $Ti^{3+}$ states. These observations in turn contribute towards the observed p-type behavior for Ir-doped BTO.

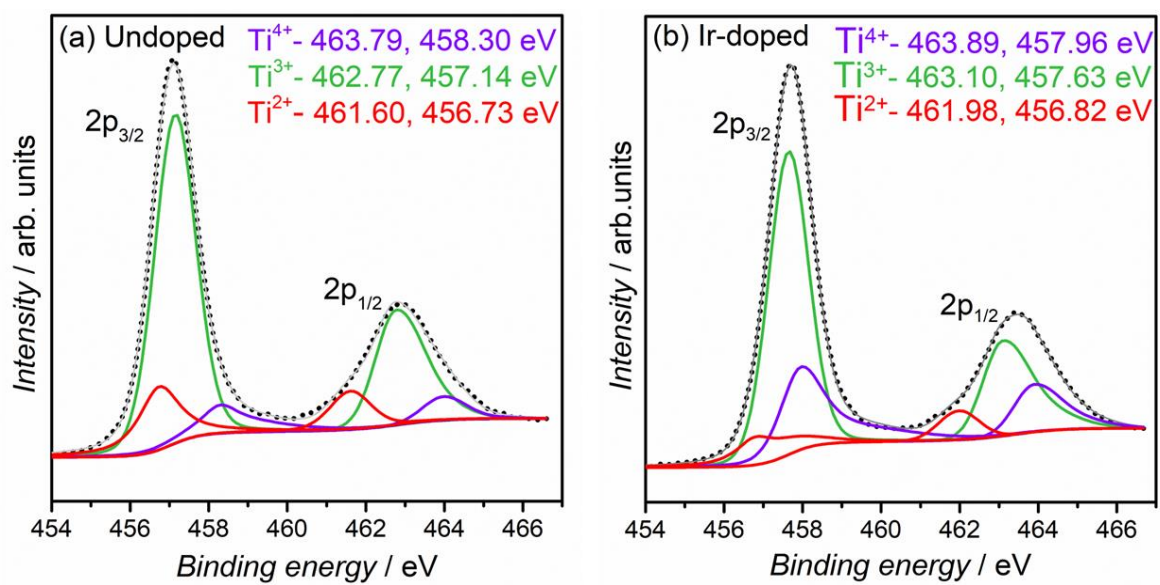

**Figure 5**. High-resolution XPS Ti 2p core level spectrum of a) undoped and b) Ir-doped BTO.

XPS data confirmed a decrease in the $Ti^{3+}$ states after Ir doping. By considering stoichiometry, the $Ti^{3+}$ and oxygen vacancies are mutually correlated. If the $Ti^{3+}$ states concentration is lowered, it naturally accompanies a decrease in oxygen vacancies. However, XPS study of oxygen vacancies does not offer quantitative information on the different types of oxygen vacancies, which is rather complex. Hence, a spin-sensitive technique such as EPR was used to probe a reduction in the concentration of oxygen vacancies after Ir doping. The EPR spectra of undoped and Ir-doped BTO in **Figure 6** shows a sharp signal at g=2.003 corresponding to oxygen vacancies.[58,59] Compared to undoped, the signal intensity is smaller for Ir-doped BTO, which indicates a decrement in the oxygen vacancy concentration. This notion is further supported by a decrement in the corresponding O 1s peak(s) area ($O_2+O_3$, corresponding to oxygen vacancies) as depicted in Figure S3 a and b. Thus, a decrease in $Ti^{3+}$ concentration and also the mutually correlated oxygen vacancy decrease upon Ir-doping is confirmed by XPS and EPR analysis, respectively.

In short, collective data from Section 3.1 revealed that Ir-doping occurs at the Ti-site of BTO host lattice. Ti exists predominantly as $Ti^{3+}$ in undoped BTO, which renders a strong n-type behavior. However, Ir doping reduces the concentration of $Ti^{3+}$ (also oxygen vacancies) and increases the concentration of $Ti^{4+}$, suggesting a transition towards p-type behavior. In the next section, the mechanistic origin of doping-induced p-type behavior in Ir-BTO will be discussed.

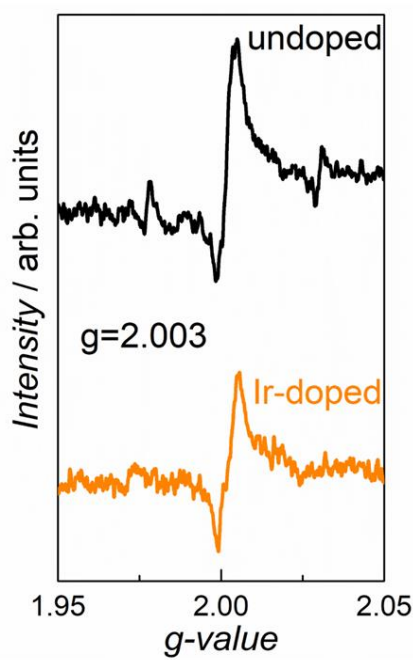

**Figure 6**. EPR spectra of undoped and Ir-doped BTO with the sharp signal at g=2.003 indicating the presence of oxygen vacancies.

**2.2. Origin of iridium-doping-induced visible-light absorption and p-type behavior**

The optical absorption spectra of undoped BTO and Ir-doped BTO are depicted in **Figure 7**a. The undoped BTO exhibits an absorption onset at around 390 nm, attributed to the fundamental transition from the VB comprising of filled O 2p orbitals to the CB consisting of Ti 3d orbitals, indicated as $T_1$ in Figure 7c.[60] Earlier reports on Ir-doped $SrTiO_3$ redshifted the optical absorption by introducing inter-band energy levels compared to its undoped counterpart.[51,61] In BTO, Ir-doping exhibits substantial redshift of optical absorption up to 600 nm, as shown in Figure 7a which can be ascribed to the formation of Ir-related in-gap energy levels. Therefore, Ir-doped BTO can harvest a wider part of the solar spectra ($\lambda \leq 600$ nm) in comparison to its undoped counterpart, which can absorb only in the UV region.

As revealed by XPS analysis, Ir species existed as $Ir^{3+}$ and $Ir^{4+}$ in the Ir-doped BTO. Ir in its $Ir^{4+}$ state has partially occupied $d^5$ electronic configuration (Ir 5d $t_{2g}^5 e_g^0$). Thus, electron transition from $Ir^{4+}$ levels to CB Ti 3d orbitals is unlikely. However, electron transition from VB to partially occupied $Ir^{4+}$ levels is feasible. On the other hand, eliciting electron transition from completely occupied $Ir^{3+}$ (Ir 5d $t_{2g}^6 e_g^0$) levels to the CB is also possible. Therefore, the absorption band in the region 400-600 nm is attributed to either the electron transition from VB to the partially occupied $Ir^{4+}$ levels, denoted as $T_2$ in Figure 7c or from the completely occupied $Ir^{3+}$ levels to the CB. However, overlapping contributions make it difficult to distinguish T2 from T3 in the optical absorption spectra.

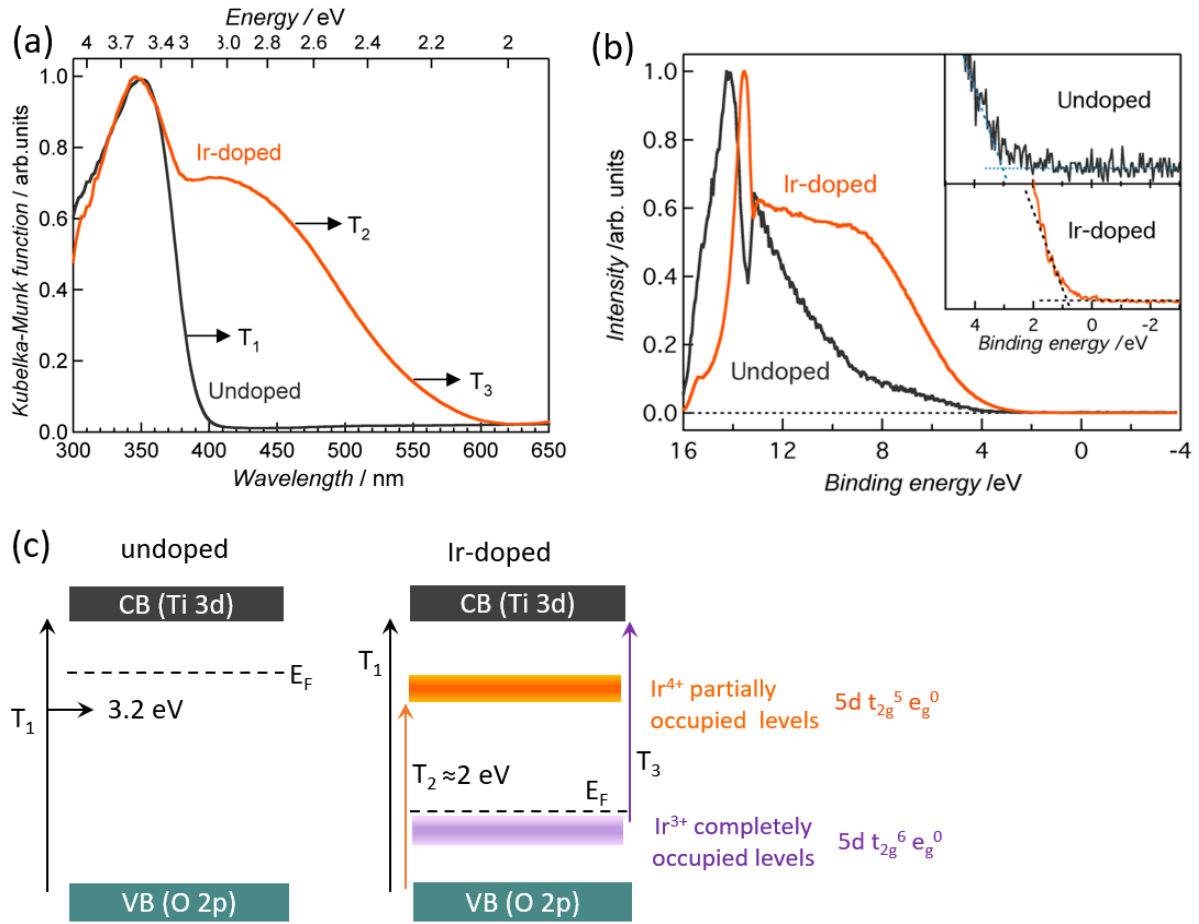

**Figure 7**. a) Diffuse reflection spectra from ultraviolet to the visible region. (b) VB-XPS spectra c) Proposed band energy diagram of undoped and Ir-doped BTO.

The effect of Ir-doping on the Fermi level position in undoped BTO and Ir-doped BTO was studied by conducting VB-XPS analysis. The shift in the VB onset towards higher (or lower) binding energy with respect to the undoped sample provides information on the relative changes in the Fermi level position due to doping. Recently, VB-XPS study resulted in deducing the p-type behavior after doping $SrTiO_3$ with $Al^{3+}$.[31] Figure 7b depicts the VB-spectra of undoped, and Ir-doped BTO with onset zoomed in the inset. The VB spectral onset of undoped BTO is observed at ≈2.2 eV, indicating that the Fermi level is positioned above the VB edge. Considering the optical band gap of 3.2 eV for the undoped BTO, it can be inferred that the Fermi level is relatively closer to the CB, in agreement with the n-type nature of undoped BTO.[62,63] For the Ir-doped BTO, VB onset is at ≈0.8 eV, indicating a significant shift of the Fermi level towards the VB edge, compared to undoped. This relative Fermi level down-shift after Ir-doping indicates a pronounced p-type behavior.

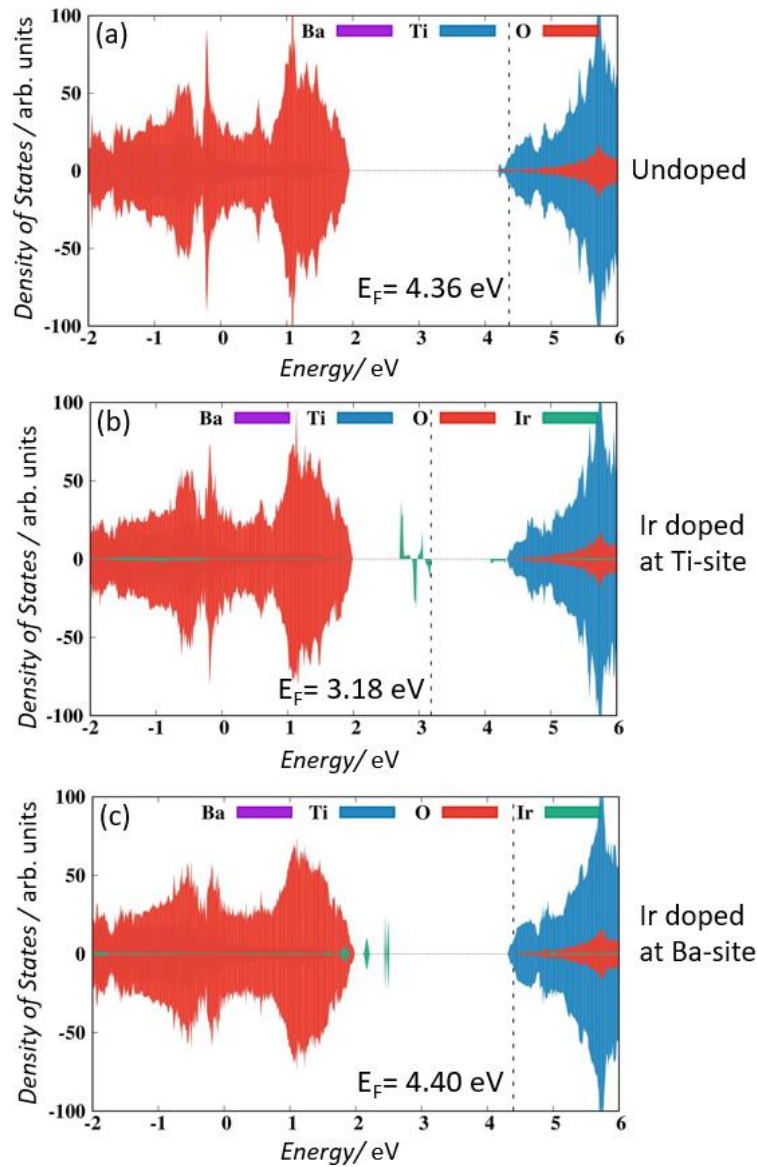

**Figure 8**. DOS of a) undoped BTO b) Ir-doped BTO with Ir at Ti-site and c) DOS of Ir-doped BTO with Ir at Ba-site.

The n- to p-type transition on Ir-doping is further corroborated by conducting photoelectrochemical measurements. If Ir-doping induces p-type behavior, a cathodic response in the photocurrent is expected. Indeed, noticing a clear photocathodic current upon light illumination further corroborates the p-type behavior in Ir-doped BTO (Figure S5). A similar approach is earlier used by Kudo et al.[32] and Maeda[33] to ascribe p-type behavior to perovskite oxides. Besides, Ir-doped BTO shows a visible-light ($\lambda$= 437 nm) responsive photocurrent, corroborating the successful role of Ir doping in extending the optical response of the BTO towards the visible region. Thus, VB-XPS data and photocathodic current infer p-type behavior to Ir-doped BTO.

The plausible mechanisms for the n- to p-type transition in BTO due to Ir-doping are discussed as follows. As already discussed using equations 1 and 2, the n-type nature is ascribed to the $Ti^{3+}$ donor levels, which is also confirmed for undoped BTO by XPS analysis (Figure 5a). Besides, XPS data of Ir-doping at Ti sites of BTO (Table 1) showed a factor of 2.6 times increase in the $Ti^{4+}$ concentration upon Ir-doping. Concurrently, a decrease in lower-valent $Ti^{3+}$ levels is also observed, which induces p-type behavior.

To further confirm the effect of Ir doping on the Fermi level position, computational analysis of the electronic structure is carried out. **Figure 8**a depicts the Density of States (DOS) of undoped BTO. Here, the Fermi level is located close to the CB region, confirming n-type behavior. Figure 8b presents the DOS of BTO when Ir is doped at Ti-site and the corresponding DOS of Ir 5d-orbitals in depicted in Figure S6. Results in Figure 8b show a substantial downshift (4.36 to 3.18 eV) of the Fermi level towards the VB, indicating a p-type behavior. This observation is in good agreement with the VB-XPS experimental data (Figure 7b). Note that Ir-doping at Ba-site (Figure 8c) did not result in a downward shift of the Fermi level, unlike doping at Ti-site. Therefore, it can be deduced that decreasing the concentration of $Ti^{3+}$ is essential to realize p-type behavior in BTO. This is achieved by preferential doping of Ir at the Ti-site, which is clearly evidenced in this work.

Another model to explain the p-type behavior is based on the generation of hole as the free charge carriers. The EPR and XPS data in Figure 6 and Figure 3a and 3b, respectively, demonstrated a reduction in the oxygen vacancy concentration upon Ir-doping at $Ti^{4+}$. It means that there is no additional formation of oxygen vacancies after doping. Further, the DOS of Ir 5d-orbitals in Ir-doped BTO depicted in Figure S6 indicated the presence of completely occupied and partially occupied levels below and above the Fermi level. This can be related to the coexistence of Ir as $Ir^{3+}$ and $Ir^{4+}$ as discussed in Figure 4. When $Ir^{3+}$ is doped at $Ti^{4+}$-site, it releases a hole into the system to maintain the charge neutrality (equation 3), thus rendering p-type behavior. Additionally, the reversible oxidation states of Ir between +3 and +4 in Ir-doped BTO would release holes, further contributing to p-type behavior, according to equation 4. This is in line with the observation reported for the origin of p-type behavior in Rh-doped $SrTiO_3$.[32] A decrease in the $Ti^{3+}$ donor levels concentration and hole formation explains p-type behavior in Ir-BTO. Thus, Ir-BTO can be a promising p-type perovskite-oxide with visible-light absorption can be utilized for a range of optoelectronics and photocatalytic applications. We are currently studying the solar hydrogen evolution activity of these samples and will be reported separately.

$$Ti^{4+} \rightleftharpoons Ir^{3+} + h^+ \quad (3)$$

$$Ir^{4+} \rightleftharpoons Ir^{3+} + h^+ \quad (4)$$

## 3. Conclusions

Doping Ir at Ti sites of BTO rendered a p-type semiconducting behavior. Collective results from VB-XPS analysis and photocathodic measurements, in conjunction with computational analysis of the electronic structure, confirmed the doping-induced p-type behavior. A redshift in the optical absorption from 390 to 600 nm after doping was attributed to the formation of Ir-related energy levels within the band gap. A systematic correlation of the results offered rational insight into the dopant site of occupancy in the host lattice and the importance of understanding defect chemistry to realize n- to p-type transition. Ir-doped BTO is an important addition to the very few existing p-type and visible-light absorbing perovskite oxides, which will find promising applications in optoelectronics and in a range of photocatalytic reactions.

## 4. Materials and methods

Barium nitrate (99%, Merck), titanium (IV) oxide nanopowder (NanoArc. anatase 99%, Alfa Aesar) and iridium (IV) oxide (99.9%, Aldrich) were used for synthesis without any further treatment. Undoped BaTiO$_3$ (undoped BTO) and Ir-doped BaTiO$_3$ (Ir-doped BTO) was synthesized using the solid-state reaction method. Stoichiometric amounts of Ba(NO$_3$)$_2$ (261.35 mg) and TiO$_2$ (79.9 mg) were ground using an agate mortar and pestle for 45 minutes. The mixture was transferred into an alumina boat crucible and was calcined at 900 ºC for 12 h at 10 ºC min$^{-1}$ ramp rate. The obtained product was named as undoped BTO. For Ir-doped BTO synthesis, the IrO$_2$ amount corresponding to 3 mol% doping was ground in agate mortar with the Ba(NO$_3$)$_2$ and TiO$_2$ for 45 minutes, followed by calcination at 900 ºC for 12 h at 10 ºC min$^{-1}$ ramp rate.

The powder X-ray diffraction was performed using Bruker D$_2$ Phaser using Cu-Kα radiation as the radiation source of wavelength $\lambda$=1.54 Å. The samples were mounted on the sample holder and the XRD pattern was recorded from 20º to 90º 2θ with 0.02º step size and a scan speed of 0.3 seconds per step.

Transmission electron microscopy (TEM) and scanning transmission electron microscopy (STEM) studies were performed using aberration corrected JEOL ARM 200 CF operated at 200 KeV. STEM probe size and collection semi angle range are 90 pm and 140-200 mrad, respectively. STEM-EDS mapping was performed using Oxford SDD detector. The samples for TEM / STEM studies were prepared by drop casting method. The powder samples were dissolved in iso-propyl alcohol and sonicated for 120 min. Using a dropper, required amount of sample liquid was extracted and casted on a gold coated lacey carbon grid. The drop cast sample was left overnight in air for drying.

The optical absorption of the samples was measured using PerkinElmer lambda 750 UV/Vis/NIR spectrometer in diffuse reflectance mode equipped with an integrating sphere.

Valence states and surface properties of the constituting elements were investigated by X-ray photoelectron spectroscopy (XPS) using Thermo-Scientific NEXSA spectrometer which uses Al K-Alpha (1486.6 eV) as the X-ray source. The operating pressure in the ultra-high vacuum range from $10^{-8}$ to $10^{-10}$ mbar. The XPS survey spectra was recorded at the pass energy of 200 eV with a step size of 1 eV. The XPS higher energy resolution spectra were recorded with a pass energy of 50 eV with a step size of 0.1 eV. Flood gun was used to eliminate the charging effects. The spectral charge correction was performed using carbon with C 1s peak appearing at 284.8 eV. Ultraviolet photoelectron spectroscopy (UPS) or valence band-XPS (VB-XPS) experiments were performed using spectrometer Thermo-Scientific NEXSA with the He I (21.22 eV) excitation source. The pass energy used was 2 eV with 0.050 eV step size. The operating pressure in the ultra-high vacuum range from $10^{-8}$ to $10^{-10}$ mbar. VB spectra of undoped and Ir-doped BTO was plotted by converting the counts vs kinetic energy (eV) to counts vs binding energy (eV). Kinetic energy to binding energy conversion was performed considering the equation, photon energy = binding energy + kinetic energy considering that the spectrometer is calibrated such that the Fermi energy will appear at binding energy zero.

Electron Paramagnetic spectroscopy (EPR) analysis was performed using JOEL Model JES FA200 spectrometer. X-band EPR spectral analysis was performed at 77.3 K using microwave frequency of 9.13 GHz, microwave power of 1.0 mW and modulation frequency of 100 kHz.

Electrodes for the photoelectrochemical measurements were prepared as follows. 2.0 mg of Ir-doped BTO was added to the solution containing 160 μL of ethanol, 40 μL of distilled water and 10 μL of nafion which was sonicated for 45 minutes to prepare the inks. Working electrode was prepared by drop-casting 20 μL of prepared ink on ITO (indium tin oxide) substrate of 0.25 cm$^2$ area and drying under the infrared-lamp. Photoelectrochemical measurements were recorded using Metrohm Autolab PGSTAT 204 potentiostat using the three-electrode system, platinum: counter electrode, Ag/AgCl: reference electrode, and ITO: working electrode. $K_2SO_4$ was used as the supporting electrolyte and the photoelectrochemical measurements were recorded at the scan rate of 5 mVsec$^{-1}$ with 40W Kessil KSPR160L LED of 427 nm.

The effect of Ir-doping on the electronic structure of BTO was investigated computationally using density functional theory (DFT) implemented in Vienna ab-initio Simulation Package (VASP).[38–40] Here, the electron-ion interactions are treated using projector augmented wave (PAW) method.[41] The generalized gradient approximation (GGA) with Perdew-Burke-Ernzerhof (PBE)[42] functional is used for electron exchange-correlation effects. The calculations are carried out with a cut-off energy of 500 eV. The Brillouin zone of the supercell is sampled using Monkhorst-Pack k-point grids of 3x3x3 and 7x7x7 for self-consistent and non-self-consistent field calculations, respectively.[43] The convergence criterion for total energy is 10$^{-6}$ eV and the residual force on each atom is less than 0.01 eVA$^{-1}$ in the relaxed structure. In a 3x3x3 supercell of BTO, there are 27 Ba atoms, 27 Ti atoms and 81 oxygen atoms as shown in Figure S1a. However, by virtue of the inherent oxygen-deficient nature of BTO (which is already detailed in section 1), oxygen deficiency is incorporated in both undoped and Ir-doped BTO (refer Figure S1 a and b). The density of states (DOS) of pristine BTO was calculated and it is found that the band gap is underestimated which is the norm of GGA for semiconductors. Next, a rotationally invariant approach of GGA+U[44] is employed where U values of 4 and 8 eV are used for Ti and O, respectively and these are taken from earlier report.[45] With this set of values, the band gap of BTO is 2.3 eV, less than what is experimentally reported. By taking $U_{Ti}$ and $U_O$ to be 8 and 8 eV respectively, the band gap increases to 2.9 eV, which is still lower than the reported experimental value. Since we are interested in the qualitative changes, we stick to the commonly used $U_{Ti}$ = 4 eV and $U_O$ = 8 eV and for Ir, an U value of 2 eV are considered.

**Supporting Information**
Supporting Information is available from the author.

**Acknowledgements**
SC and DHKM acknowledge Technology Mission Division (Energy, Water & all Others), Department of Science & Technology, Ministry of Science & Technology, Government of India, reference Number DST/TMD/IC-MAP/2K20/02, project titled as Integrated Clean

Energy Material Acceleration Platform (IC-MAP) on bioenergy and hydrogen. P. S. S. R. K acknowledge financial support from A*STAR, Singapore under the Structural Metals and Alloys Program (Grant No.: A18B1b0061). DHKM would like to thank Dr. Nagaraja K K and Dr. Mallikarjun Bhavanari for their comments on the manuscript.

**Conflict of Interest**

The authors declare no conflict of interest.